\documentclass[aps,prb,twocolumn,groupedaddress,floatfix,showpacs]{revtex4}
\usepackage{graphicx}
\usepackage{amsmath}

\begin{document}

\title{Non-local Andreev transport through an interacting quantum dot}
\author{David Futterer$^1$} 
\author{Michele Governale$^1$}
\author{Marco G. Pala$^2$}
\author{J\"urgen K\"onig$^{1,3}$}

\affiliation{
$^1$Institut f\"ur Theoretische Physik III,
Ruhr-Universit\"at Bochum, D-44780 Bochum, Germany\\
$^2$IMEP-LAHC-MINATEC (UMR CNRS/INPG/UJF 5130), 38016 Grenoble, France\\
$^3$Theoretische Physik, Universit\"at Duisburg-Essen, 47048 Duisburg, Germany
}

\date{\today}
\begin{abstract}
We investigate sub-gap transport through a single-level quantum dot tunnel coupled to one superconducting and two normal-conducting leads.
Despite the tendency of a large charging energy to suppress the equilibrium proximity effect,
a finite Andreev current through the dot can be achieved in non-equilibrium situations.
We propose two schemes to identify non-local Andreev transport.
In one of them, the presence of strong Coulomb interaction leads to negative values of the non-local conductance as a clear signal of non-local Andreev transport.  
\end{abstract}

\maketitle 

\section{Introduction}
 
The entanglement of electrons in Cooper pairs of a superconductor can generate non-local transport effects.
A prominent example is \textit{Crossed Andreev Reflection} (CAR) at the contact between a superconductor and two normal leads: 
there the two electrons with opposite spins and symmetric energies with respect to the Fermi level of the superconductor, that are transferred through the normal-superconductor interface via Andreev 
reflection,\cite{andreev64} originate from or end up in different normal leads.
This is a non-local transport phenomenon which has been extensively studied both 
theoretically\cite{falci01,lesovik01,yamashita03,sanchez03,melin04,morten06,brinkmann06,kalenkov07,golubev07,kalenkov07_2} and experimentally.\cite{bozhko82,benistant83,beckmann04,russo05,cadden-zimansky06}

The main problem to identify the non-locality of CAR in a transport measurement is to separate it from other transport channels.
In this paper, we study non-local Andreev transport through a single-level quantum dot.
The quantum-dot level energy can be tuned by a gate voltage, which opens the possibility to control 
the Andreev transport channels. 
At first glance, strong Coulomb interaction in the quantum dot seems to be counterproductive:
the formation of a finite equilibrium superconducting pair amplitude is suppressed since a large charging energy prevents the equilibrium state to be a coherent superposition of dot states with particle numbers differing by two, and Cooper pairs can be transferred through the dot by higher-order tunneling processes (cotunneling) only. 
On the other hand, a finite \textit{non-equilibrium} pair amplitude in the dot can be achieved with a bias voltage.\cite{pala07,governale08} 
A large charging energy provides even the key ingredient for identifying non-local Andreev transport in one of the schemes we propose.

\begin{figure}
\includegraphics[width=2.6in]{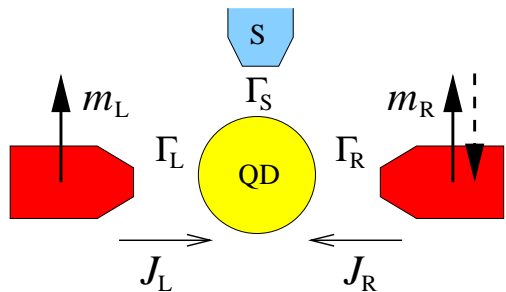}
\caption{(color online)
Setup: a quantum dot is tunnel coupled to one superconducting and two normal-conducting leads. The latter may be ferromagnetic with magnetization
directions $\mathbf{\hat{m}_{\rm{L}}}$ and $\mathbf{\hat{m}_{\rm{R}}} = \pm \mathbf{\hat{m}_{\rm{L}}}$.
\label{setup}}
\end{figure}

Andreev reflection through quantum dots, a problem which combines Coulomb interaction, superconducting correlations and non-equilibrium, has been extensively studied 
theoretically.\cite{fazio98,fazio99,kang98,schwab99,clerk00,cuevas01,pala07,governale08}
Here, we apply the diagrammatic real-time transport theory of 
Ref.~\onlinecite{pala07,governale08}.
The relevance from the experimental point of view is proven by the recent success in coupling superconductors to quantum dots in either a carbon nanotube\cite{buitelaar02,cleuziou06,jarillo-herrero06,jorgensen06} or a 
semiconductor nanowire.\cite{van_dam06,sand-jespersen07,buizert07}
In particular, we propose to investigate non-local effects in Andreev transport through a 
single-level quantum dot with one superconducting and two normal-conducting leads, which may be (i) ferromagnetic with collinear magnetization or (ii) non-magnetic (see Fig. \ref{setup}).
Thereby we identify non-locality either (i) by the dependence of the Andreev current on the relative orientation of the two ferromagnets that are biased with the same voltage, or (ii) by the response of the current in one normal lead to the applied voltage in the other one.

\section{Model}
 
The Hamiltonian of the system is $H=H_{\rm{dot}}+\sum_{\eta}(H_{\eta}+H_{\rm{tunn},\eta})$. 
The label $\eta= \rm{L},\rm{R},\rm{S}$ corresponds to the left, right and the superconducting lead, respectively.
The dot is described by the Anderson-impurity model:
\begin{equation} 
H_{\rm{dot}}=\sum_{\sigma} \epsilon d_{\sigma}^{\dagger} d_{\sigma} + U n_{\uparrow} 
n_{\downarrow},
\end{equation}
where $d_{\sigma}^{(\dagger)}$ is the annihilation (creation) operator for a dot electron with spin $\sigma$, $n_{\sigma}=d_{\sigma}^{\dagger} d_{\sigma} $ is the corresponding number operator, $\epsilon$ the energy of the spin-degenerate single-particle level, and $U$ the on-site Coulomb repulsion. 
The Hamiltonian of the lead $\eta$ reads
\begin{equation}
H_{\eta}=\sum_{k \sigma} \epsilon_{\eta k\sigma}
c_{\eta k \sigma}^\dagger c_{\eta k \sigma}- g_\eta \sum_{k,k'} c_{\eta k \uparrow}^\dagger 
c_{\eta -k \downarrow}^\dagger c_{\eta -k' \downarrow} 
c_{\eta k' \uparrow}, 
\end{equation}
where the single-particle energies $\epsilon_{\eta k\sigma}$ are spin dependent in the case of ferromagnetic leads (with the quantization axis for $\sigma$ being along the magnetization direction of the left lead), and the BCS pairing-interaction strength $g_\eta$ is non-vanishing only for $\eta={\rm S}$.
The lead-electron operators are $c_{\eta k \sigma}$ and $c_{\eta k \sigma}^\dagger$.
We treat the superconductor on a mean-field level, which introduces the notion of fermionic quasiparticles and a Cooper-pair condensate. 
The coupling between the dot and the leads is taken into account by the 
tunneling Hamiltonians
\begin{equation}
  H_{\rm{tunn},\eta}= 
  V_{\eta} \sum_{k \sigma} \left( c_{\eta k \sigma}^\dagger d_\sigma +
  {\rm H.c.} \right), 
\label{htun}
\end{equation}
where for simplicity the tunnel matrix 
elements $V_{\eta}$ are considered to be spin and wavevector independent.

The spin-resolved tunnel-coupling strengths are defined as
$\Gamma_{\eta\sigma}=2 \pi |V_{\eta}|^2 \sum_k \delta (\omega - \epsilon_{k\sigma})=2\pi |V_\eta|^2 \rho_{\eta\sigma}$,  with $\rho_{\eta\sigma}$ being  
the density of states of the spin species $\sigma$ in lead $\eta$, which we assume to be constant.
We also define the mean level-broadening $\Gamma_\eta=\frac{1}{2}\sum_{\sigma}\Gamma_{\eta\sigma}$.
The spin polarization of lead $\eta$ is defined as
$p_\eta=(\rho_{\eta+}-\rho_{\eta-})/(\rho_{\eta+}+\rho_{\eta-})$, 
with $+(-)$ denoting the majority(minority) spins.

\section{Method}
 
We integrate out the 
leads' degrees of freedom to obtain an effective description of the quantum dot,
whose Hilbert space is spanned by the four basis states 
$|\chi\rangle \in \{ | 0 \rangle, | \uparrow \rangle,
| \downarrow \rangle, | {\rm{D}} \rangle\equiv d^{\dagger}_{\uparrow} 
d^{\dagger}_{\downarrow}|0\rangle \}$, with energies
$E_0$, $E_\uparrow=E_\downarrow$, and $E_{\rm{D}}$, corresponding to an empty, singly and doubly occupied dot.
It is useful to define the detuning $\delta=E_{\rm{D}}-E_0=2\epsilon+U$.
The dot is described by its reduced density matrix, $\rho_{\rm{red}}$, whose matrix elements 
are $P_{\chi_2}^{\chi_1} \equiv\langle \chi_1|\rho_{\rm{red}}
|\chi_2\rangle$.
The superconducting proximity effect induces a finite pair amplitude on the dot, expressed by the 
off-diagonal matrix element $P_{0}^{\rm D} = (P_{\rm D}^{0})^*$.

In the stationary limit, the elements of the reduced density matrix obey the generalized master equation 
\begin{equation}
i(E_{\chi_1}-E_{\chi_2}) P^{\chi_1}_{\chi_2}=\sum_{\chi_1' \chi_2'} W^{\chi_1 \chi_1'}_{\chi_2 \chi_2'}P^{\chi_1'}_{\chi_2'} \,
\end{equation}
with generalized rates $ W^{\chi_1 \chi_1'}_{\chi_2 \chi_2'}$, that can be computed
by means of a real-time diagrammatic technique.\cite{koenig96,koenig96_2,koenig03,braun04,pala07,governale08}  
The stationary current out of lead $\eta$ can be expressed as 
\begin{equation}
  J_{\eta} = -\frac{e}{\hbar} \sum_{\chi \chi_1' \chi_2'}  
  W_{\chi \chi_2'}^{\chi \chi_1' \eta} 
  P_{\chi_2'}^{\chi_1'},
\label{current3}
\end{equation}
where $W_{\chi \chi_2'}^{\chi \chi_1' \eta} \equiv
\sum_s s W_{\chi \chi_2'}^{\chi \chi_1' s \eta}$, and 
$W_{\chi \chi_2'}^{\chi \chi_1' s \eta}$ is the sum of all generalized rates that
describe transitions in which in total $s$ electrons are removed from lead 
$\eta$.

\begin{figure}
\includegraphics[width=3.1in]{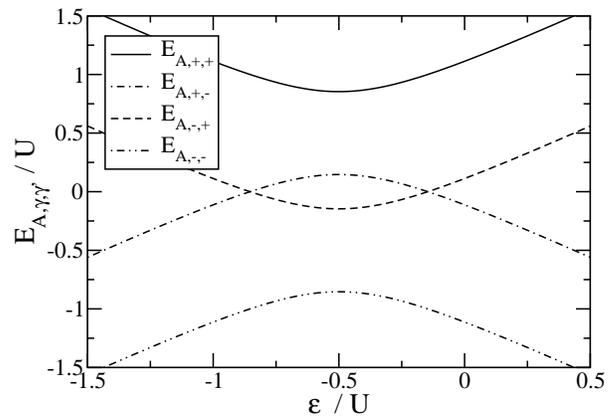}
\caption{ 
Andreev bound-state energies as a function of the level position $\epsilon$, with $\Gamma_{\rm {S}}/U=0.2$ and $\Gamma_{\rm{N}}=0$.
\label{boundstates}}
\end{figure}

The diagrammatic rules to compute the diagrams contributing to the rates in the presence of superconducting and ferromagnetic leads are given in Refs.~\onlinecite{governale08} and \onlinecite{koenig03,braun04}, respectively.  
In the following, we assume small and equal tunnel-coupling strengths $\Gamma_{\rm{L}} =\Gamma_{\rm{R}}  \equiv \Gamma_{\rm{N}} < T$ to the left and right lead, which we keep up to first order
in the calculation of the current.
The chemical potential of the superconductor is chosen as the reference for the energies, i.e. 
$\mu_{\rm{S}}=0$.
To study subgap transport, we consider the limit of a large superconducting order parameter 
$\Delta \rightarrow \infty$, which we choose to be real.
In this case, all orders in the tunnel-coupling strength with the superconductor $\Gamma_{\rm{S}}$ can be resummed exactly.\cite{governale08,rozhkov00}
In the absence of a superconducting lead, the excitation energies of the dot, $\epsilon$ and $\epsilon+U$, 
are split by the charging energy $U$.
Due to the tunnel coupling to a superconductor the particle and hole sector of the Hilbert space are mixed,
leading to four Andreev bound-state energies, defined as poles of the retarded Green's function of the dot for $\Gamma_{\rm N} =0$:
\begin{equation}
\label{ABS}
E_{\rm{A}, \gamma', \gamma} = \gamma' \frac{U}{2} + \gamma \sqrt{\left( \epsilon + \frac{U}{2} \right)^{2} + \frac{\Gamma_{\rm S}^{2}}{4}},
\end{equation}
with $\gamma', \gamma = \pm 1$ (see Fig. \ref{boundstates}). 
In the following, we also make use of the 
definition $\epsilon_{\rm A} \equiv \sqrt{(\epsilon + \frac{U}{2})^{2} + \frac{\Gamma_{\rm S}^{2}}{4}}$.
The Andreev transport is largest for small detuning $|\delta|$. 
We consider the regime $|\delta| < \sqrt{U^2 - \Gamma_{\rm{S}}^2}$, for which the inequality 
$E_{\rm{A},+,+} > E_{\rm{A},+,-} > 0 > E_{\rm{A},-,+} > E_{\rm{A},-,-}$ holds.

The Andreev channel supporting transport from a normal lead to a superconductor through a quantum dot can be switched between different states by an applied bias voltage 
$\mu_{\rm{N}}$.\cite{governale08}
In the following we consider $\mu_{\rm{N}} > 0$ (the case $\mu_{\rm{N}} < 0$ is obtained from the symmetry transformation $\mu_{\rm{N}} \rightarrow -\mu_{\rm{N}}$, $\delta \rightarrow - \delta$, and 
$J_{\rm{N}} \rightarrow -J_{\rm{N}}$).
The different states are characterized by how they influence and probe the state of the quantum dot.

For small bias voltage, $E_{\rm{A},+,-} > \mu_{\rm{N}}$, all rates (to first order in $\Gamma_{\rm{N}}$) involving an electron transfer from or to the normal lead are not affected by the superconductor. 
The current can be written in the simple form $J_{\rm{N}} = \frac{2e}{\hbar} \Gamma_{\rm{N}} \left( 1 + Q/e \right)$, independent of the pair amplitude $P_{0}^{\rm D}$, 
where $Q/(-e) = \langle \sum_{\sigma} d_{\sigma}^{\dagger}d_{\sigma} \rangle =P_{\uparrow}+P_{\downarrow}+2P_{\rm{D}}$ (with $-2e \le Q \le 0$) is the average quantum-dot charge. The latter can be affected by the proximity effect in the quantum dot. 
In the stationary limit, the dot is singly occupied, $Q=-e$, and the Andreev channel is Coulomb blocked, $J_{\rm{N}} = 0$. A current can only flow for $Q \neq -e$, which. e.g., could be achieved by attaching a voltage-biased third lead.
For large bias voltage, $\mu_{\rm{N}} > E_{\rm{A},+,+}$, the Andreev channel is also independent of the dot pair amplitude, with $J_{\rm{N}} = \frac{2e}{\hbar} \Gamma_{\rm{N}} \left( 1 + Q/2e \right)$.
In both cases, the sub-gap transport involves Andreev processes at the interface between quantum dot and the superconducting lead only.
As a consequence, the current is, in both cases, insensitive to the sign of the detuning $\delta$.
This is different for the regime of intermediate voltages, $E_{\rm{A},+,+} > \mu_{\rm{N}} > E_{\rm{A},+,-}$. In this case the rates involving an electron transfer to or from the normal lead do depend on $\delta$ and $\Gamma_{\rm{S}}$,  
the rates $W_{0 \sigma}^{\rm{D} \sigma}$ and 
$W_{\rm{D} \sigma}^{0 \sigma}$ describing proximization of the quantum dot are non-vanishing, and
the current in the normal lead also depends on the pair amplitude $P_{0}^{\rm D}$.
An interesting feature of this regime is that a positive detuning drives the dot to an average occupation of less than one electron, thus overcompensating the effect of the finite bias voltage.\cite{governale08}

\section{Crossed Andreev Reflection}
  
\begin{figure}
\includegraphics[width=3.1in]{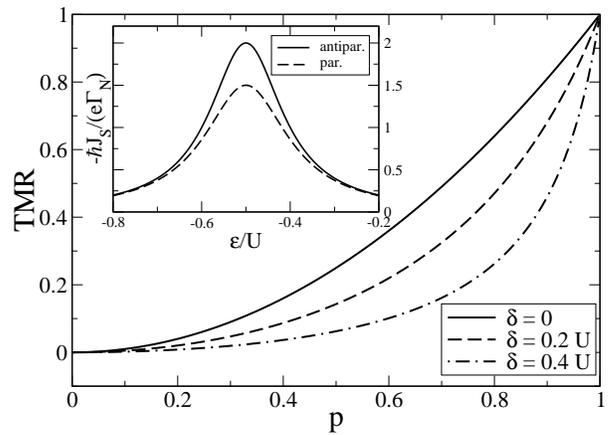}
\caption{ 
TMR as a function of the polarization $p$, for different values of the detuning $\delta=2\epsilon+U$. 
Inset: Current $J_{\rm{S}}$ in the superconducting lead as a function of 
$\epsilon$,
for polarization $p=0.5$ and parallel and antiparallel alignment of the leads' magnetizations.
For both graphs, we chose $\mu_{\rm{L}}=\mu_{\rm{R}}=\mu_{\rm{N}}=U$, $\mu_{\rm{S}}=0$,
$\Gamma_{\rm{S}}=0.2U$, $\Gamma_{\rm{L}}=\Gamma_{\rm{R}}=0.001 U$, and $k_{\rm{B}}T=0.01 U$.
\label{car1}}
\end{figure}

We first consider the case of ferromagnetic leads with equal polarization strengths 
$|p_{\rm{L}}|=|p_{\rm{R}}|=p$, kept at the same chemical potential $\mu_{\rm{N}}$ and characterized by the same Fermi distribution $f_{\rm{N}}(\epsilon)=\left[ 1 + \exp{(\epsilon - \mu_{\rm{N}})/k_{\rm{B}}T} \right]^{-1}$. 
Crossed Andreev transport
is identified by its dependence on the relative orientation of the ferromagnets, quantified by the Tunneling Magneto Resistance (TMR)  
${\rm{TMR}}\equiv (J_{\rm{S}}^{\rm{AP}}-J_{\rm{S}}^{\rm{P}})/ J_{\rm{S}}^{\rm{AP}}$, 
where $J_{\rm{S}}^{\rm{P(AP)}} = (2 e\Gamma_{\rm{S}}/\hbar) \rm{Im} P_{0}^{\rm{D}}$ is the current in the superconductor for parallel (antiparallel) alignment of the magnetizations. 
The pair amplitude $P_{0}^{\rm{D}}=\langle d_{\downarrow} d_{\uparrow} \rangle$ is calculated in the stationary limit from the master equation $P_{0}^{\rm{D}} = \sum_{\chi} W_{0\chi}^{\rm{D}\chi} P_{\chi}^{\chi} / \left[ W_{00}^{\rm{DD}} - i(2\epsilon + U) \right]$.
In the inset of Fig.~\ref{car1} we show the current in the superconducting lead for both the parallel and antiparallel alignment of the magnetizations as a function of the gate voltage for $\mu_{\rm{N}}=U$ (large-bias regime).  
The current shows a peak around zero detuning, $\epsilon=-U/2$, with a width 
given by $\Gamma_{\rm{S}}$. 
The sub-gap 
current for the parallel 
alignment 
is clearly suppressed, indicating the presence of CAR.
The TMR is plotted in Fig.~\ref{car1} as a function of the polarization strength $p$ for different values of the detuning $\delta$.
For small values of $p$
the TMR exhibits a quadratic dependence on $p$, 
\begin{equation}
	{\rm{TMR}}\approx  \frac{\left[1-\Pi_{\gamma=\pm}\left(\sum_{\gamma'=\pm}\gamma' f_{\rm{N}}	(E_{\rm{A},\gamma',\gamma})\right)\right] \Gamma_{\rm{S}}^2 p^2} {4\epsilon_{\rm{A}}^2-
	\left[\sum_{\gamma,\gamma'=\pm} \gamma \gamma'(\epsilon+\frac{U}{2}+\epsilon_{\rm{A}}) 	f_{\rm{N}}(E_{\rm{A},\gamma',\gamma})\right]^2}, 
\end{equation}
which simplifies in the regime of large bias voltage to 
${\rm{TMR}} \approx p^2$ for $|\delta| \ll \Gamma_{\rm{S}}$ and
${\rm{TMR}} \approx p^2 \Gamma_{\rm{S}}^2 / \delta^2$ for $|\delta| \gg \Gamma_{\rm{S}}$, and in the regime of intermediate (positive) bias voltage to
${\rm{TMR}} \approx 4p^2/3$ for $|\delta| \ll \Gamma_{\rm{S}}$,
${\rm{TMR}} \approx p^2 \Gamma_{\rm{S}}^2/\delta^2$ for $-\delta \gg \Gamma_{\rm{S}}$, and
${\rm{TMR}} \approx 2p^2$ for $\delta \gg \Gamma_{\rm{S}}$.

\begin{figure}
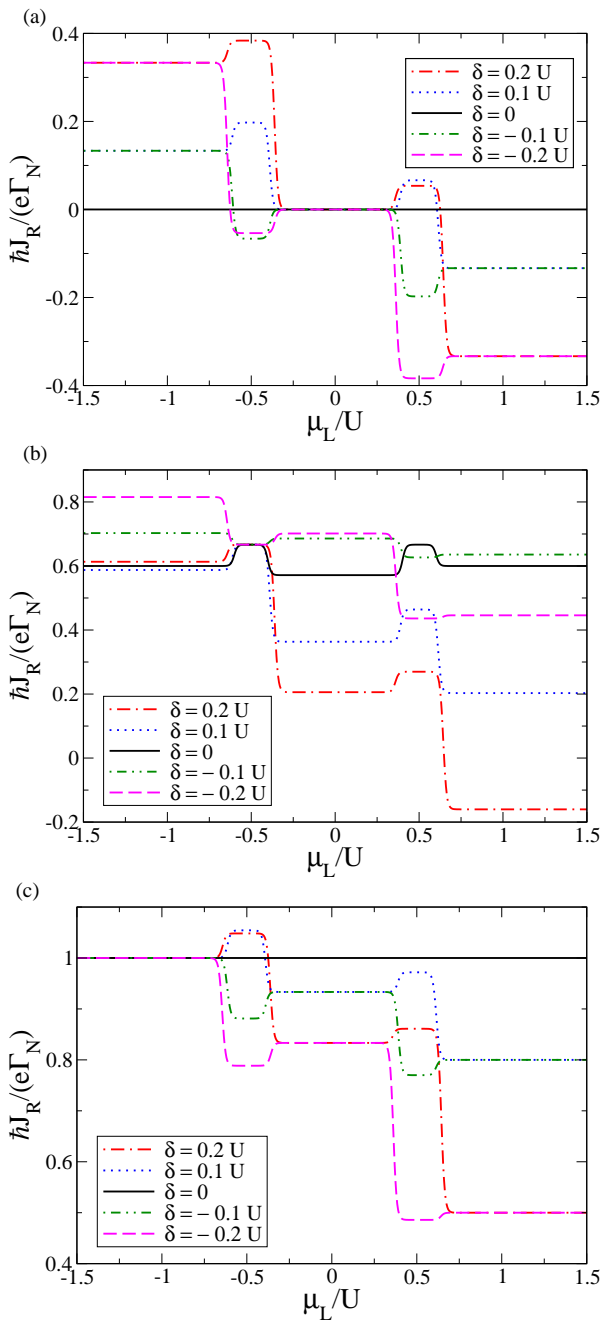

\includegraphics[width=3.1in]{fig4a.eps}\\
\includegraphics[width=3.1in]{fig4b.eps}\\
\includegraphics[width=3.1in]{fig4c.eps}
\caption{(color online) Current $J_{\rm{R}}$ in the right lead as a function of the chemical potential $\mu_{\rm{L}}$ of the left lead for different values of the detuning $\delta$ and 
$\Gamma_{\rm{S}}=0.2 U$, $\Gamma_{\rm{L}}=\Gamma_{\rm{R}}=0.001U$,    $p_{\rm{L}}=p_{\rm{R}}=0$, $k_{\rm{B}}T = 0.01 U$, and 
(a) $\mu_{\rm{R}}=0$ (small bias),
(b) $\mu_{\rm{R}}=0.5 U$ (intermediate bias), and
(c) $\mu_{\rm{R}}=U$ (large bias).
\label{rev1}}
\end{figure}

\section{Negative non-local conductance}

We now look for non-local effects in Andreev transport for non-magnetic leads,
$p_{\rm{L}}=p_{\rm{R}}=0$, by studying the current in the right lead, $J_{\rm{R}}$, as a function of the voltage applied to the left lead $\mu_{\rm{L}}$.
For a three-terminal device, we define a non-local conductance, 
$G^{\rm nl} \equiv J_{\rm{R}} / (\mu_{\rm{S}} - \mu_{\rm{L}})$, as the current response in 
the right lead to a voltage bias between superconductor and left lead.
In particular, we will consider the non-local differential conductance
$G^{\rm nl}_{\rm diff} \equiv -\partial J_{\rm{R}} / \partial \mu_{\rm{L}}$.
Direct transport between the two normal leads contributes with a positive sign to $G^{\rm nl}_{\rm diff}$.
Non-local transport channels such as CAR may contribute with a negative sign.
For a large class of such three-terminal devices, however, it has been shown\cite{morten07} 
that the sum of all contributions to $G^{\rm nl}_{\rm diff}$ remains positive.
In contrast, we find for our system regimes with negative values of $G^{\rm nl}_{\rm diff}$ and, even more striking, negative values of $G^{\rm nl}$.

In Fig.~\ref{rev1}a, b, and c, we show the current in the right lead as a function of the electrochemical potential in the left lead
for the three cases of a small, intermediate and large voltage $\mu_{\rm{R}}$ applied to the current probe, respectively. 
Several features of these current-voltage characteristics indicate the strong coupling of the
quantum dot to a superconducting lead.
First, there are four steps (instead of two) associated with the four Andreev bound-state energies. 
Second, the height of many of the plateaus is sensitive to the detuning $\delta$.
In the cases displayed in Figs.~\ref{rev1}a and c, the plateau height of the current, given by $J_{\rm{R}} = 2e \Gamma_{\rm{N}} \left( 1 + Q/e \right)$ and $J_{\rm{R}} = 2e \Gamma_{\rm{N}} \left( 1 + Q/2e \right)$, respectively, directly reflects the average quantum-dot charge, which is strongly influenced by the proximity effect.\cite{blatter07,governale08}
For $\delta=0$, the proximity effect is maximal, with $Q=-e$ for all values of the bias voltages, which leads
to $J_{\rm{R}}=0$ and $J_{\rm{R}}=e \Gamma_{\rm{N}}$ in Fig.~\ref{rev1}a and c, respectively.
With increasing $|\delta|$, the proximity effect decreases, and the current approaches the value expected
in the absence of the superconducting lead. 
Third, the striking feature indicating non-local Andreev transport is the non-monotonic dependence of $J_{\rm{R}}$ on $\mu_{\rm{L}}$, i.e., the appearance of a negative non-local differential conductance, $G^{\rm nl}_{\rm diff} < 0$. 
Even more remarkable is for $\mu_{\rm{R}}=0$ the negative non-local conductance $G^{\rm nl} < 0$
that occurs for positive/negative detuning $\delta $ at positive/negative $\mu_{\rm{L}}$ in the 
intermediate-bias regime.
To understand this behavior, we realize that in this regime there are combined Andreev processes that involve
both the interfaces from the quantum dot to the superconductor and the left lead, while there are no Andreev processes involving electron transfer between quantum dot and right lead. The intermediate-voltage Andreev transport between left lead and superconductor yields an average dot charge that is determined by the dot level's position relative to the chemical potential of the superconductor rather than that of the normal lead: for positive (negative) detuning the probability of double occupation decreases (increases), and the average occupation of the dot is smaller (larger) than one.
This deviation from single occupancy is probed by the right lead. Changing the sign of detuning leads to a sign change in the current measured in the right lead.

We remark that the negative non-local conductance is not due to CAR.
In fact, making the normal-conducting leads ferromagnetic suppresses the negative non-local conductance, independent of whether the ferromagnets are aligned parallel or antiparallel, since
a finite spin accumulation on the quantum dot reduces the dot pair amplitude.
For CAR an enhanced (reduced) effect would be expected for the antiparallel (parallel) alignment. 
The effect that we predict is rather a consequence of combined Andreev processes between left lead and superconductor.
The negative non-local conductance can only be probed because of a large charging energy that prohibits direct transport between the normal-conducting leads.

\section{Conclusions}
 
We investigated non-local Andreev transport 
through an interacting quantum dot in a three-terminal setup with one superconducting 
and two normal-conducting leads.
We considered two different biasing schemes. 
In the first one, the 
normal-conducting leads are ferromagnetic with collinear magnetizations and they are kept at the same chemical potential. 
The key results for this case is that 
CAR occurs due to the non-equilibrium proximity effect in the dot and it is characterized by a finite TMR. 
In the second scheme, the two 
normal-conducting leads are non magnetic, 
and the response of the current in one lead to the voltage applied to the other one 
is studied. 
In that case, non-local Andreev transport is identified by negative values of the non-local 
differential conductance.
Even more strikingly, we find regimes with negative values of the full non-local conductance.
The virtue of employing quantum dots lies, first, in the possibility to tune the Andreev channels
by a gate voltage and, second, in the presence of a large charging energy which generates specific transport regimes characterized by a negative non-local conductance.
Both aspects are advantageous for a clear identification of non-local Andreev transport.

\acknowledgements
We acknowledge useful discussions with W. Belzig, M. Eschrig, A. Zaikin, 
and financial support from DFG via SFB 491.


\end{document}